# Soft elastic constants from phonon spectroscopy in hole-doped Ba$_{1-x}$(K,Na)$_x$Fe$_2$As$_2$ and Sr$_{1-x}$Na$_x$Fe$_2$As$_2$

M. Kauth[1], S. Rosenkranz[2], A. H. Said[3], K. M. Taddei[2], Th. Wolf[1], F. Weber[1,4]

[1]*Institute for Solid State Physics, Karlsruhe Institute of Technology, D-76021 Karlsruhe, Germany*
[2]*Materials Science Division, Argonne National Laboratory, Lemont, Illinois, 60439, USA*
[3]*Advanced Photon Source, Argonne National Laboratory, Lemont, Illinois, 60439, USA*
[4]*Institute for Quantum Materials and Technologies, Karlsruhe Institute of Technology, D-76021 Karlsruhe, Germany*

We report inelastic x-ray scattering measurements of the in-plane polarized transverse acoustic phonon mode propagating along $q \parallel [100]$ in various hole-doped compounds belonging to the 122 family of iron-based superconductors. The slope of the dispersion of this phonon mode is proportional to the square root of the shear modulus $C_{66}$ in the $q \rightarrow 0$ limit and, hence, sensitive to the tetragonal-to-orthorhombic structural phase transition occurring in these compounds. In contrast to a recent report for Ba(Fe$_{0.94}$Co$_{0.06}$)$_2$As$_2$ [F. Weber *et al.*, Phys. Rev. B **98**, 014516 (2018)], we find qualitative agreement between values of $C_{66}$ deduced from our experiments and those derived from measurements of the Young's modulus in Ba$_{1-x}$(K,Na)$_x$Fe$_2$As$_2$ at optimal doping. These results provide an upper limit of about 50 Å for the nematic correlation length for the investigated optimally hole-doped compounds. Furthermore, we also studied compounds at lower doping levels exhibiting the orthorhombic magnetic phase, where $C_{66}$ is not accessible by volume probes, as well as the C$_4$ tetragonal magnetic phase.

## I. Introduction

Superconductivity in the 122-family of iron-based superconductors emerges when a stripe antiferromagnetic spin-density-wave (SDW) ground state is suppressed either by doping or pressure [1]. The magnetic ordering is accompanied by a transition from a tetragonal ($C_4$) to orthorhombic ($C_2$) structure because of magnetoelastic coupling [2]. While the magnetic and structural phase transitions occur simultaneously in the 122 parent compounds (T$_s$ = T$_N$ = T$_{s,N}$), the structural transition can precede the magnetic one, e.g. in Ba(Fe$_{1-x}$Co$_x$)$_2$As$_2$ [3], or continue to occur simultaneously, e.g. in Ba$_{1-x}$K$_x$Fe$_2$As$_2$ [4,5]. While it is generally assumed that the same type of fluctuations, termed nematic, are responsible for both magnetic and structural/nematic phase transitions in these compounds, the nematic phase without long-range magnetic order is only observed at temperatures T$_N$ < T < T$_s$ in the former type of iron pnictides [3,6] and chalcogenides [7].

Nematic order triggers orthorhombic order [8]. Hence, the nematic order parameter $\varphi$ has zero wave vector ($q = 0$) and is proportional to the shear distortion $\epsilon_{66}$ [9,10] due to the nemato-elastic coupling $\lambda \varphi \epsilon_{66}$ in the free energy. In the para-nematic phase with tetragonal symmetry at T > T$_s$, nematic fluctuations soften the shear modulus $C_{66}$. The same coupling then yields the general relation

$$C_{66}(q) = \frac{C_{66}^0}{1+\frac{\lambda^2}{C_{66}^0}\chi_{nem}(q)} \quad (1)$$

between the elastic constant, the bare elastic modulus $C_{66}^0$ without nematicity, and the nematic susceptibility $\chi_{nem}$. Eq.(1) was derived in Ref. [9] for $q = 0$. $C_{66} \equiv C_{66}(q = 0)$ has been measured either directly by resonant ultrasound or indirectly via the Young's modulus $Y_{110}$ in three-point bending setups [9,11,12].

Recently, some of us reported inelastic neutron scattering measurements in Ba(Fe$_{0.94}$Co$_{0.06}$)$_2$As$_2$ investigating the in-plane polarized transverse acoustic (TA) phonon mode propagating along the [010] direction (tetragonal notation) at small wavevectors [13]. In the $q = 0$ limit, the slope of this phonon mode is proportional to the square root of the shear modulus, $\sqrt{C_{66}}$. Hence, a softening/hardening of C$_{66}$ results in a phonon softening/hardening at finite wavevectors observable in phonon spectroscopy. The advantage of inelastic neutron scattering is that it yields reliable results also in the orthorhombic phase without the need for uniaxial strain or pressure for detwinning. Without detwinning bulk probes such as ultrasound [12] or a three-point bending setup in a capacitance dilatometer [11,14] cannot be used.

Here, we report measurements of the in-plane polarized TA phonon mode similar to those in [13], but in several hole-doped iron-based superconductors, Ba$_{1-x}$(K,Na)$_x$Fe$_2$As$_2$ and Sr$_{1-x}$Na$_x$Fe$_2$As$_2$. Measurements were done using inelastic x-ray scattering (IXS) in order to be able to investigate small specimen unsuited for inelastic neutron scattering. Furthermore, the high intensities of low-energy acoustic phonons allowed us to use a setup with very good momentum resolution, enabling measurements in close vicinity to a strong Bragg reflection. After a short description of the experimental details in section II, we present our experimental observations in section III. Our report is divided in optimally doped samples, i.e., samples having a high superconducting transition temperature and not displaying any structural phase transitions, and underdoped samples featuring full magnetic-structural phase transitions at temperatures $T_{s,N}$. Two samples also exhibit the $C_4$ reentrant tetragonal magnetic phase below another structural transition at $T_1$ [15]. We compare our results with previous reports on the Young's modulus $Y_{[110]}$ [11,14] since $Y_{[110]}$ is closely related to the soft shear modulus $C_{66}$ in iron-based superconductors of the 122 family. In section IV, we put our results in the broader context of research on nematicity and end with the conclusion in section V.

## II. Experimental

The experiment was performed at the XOR 30-ID high energy-resolution inelastic x-ray scattering (HERIX) beamline [16] [17] of the Advanced Photon Source, Argonne National Laboratory. The samples were high-quality single crystals, grown by a self-flux method. Samples had typical dimensions of $1 \times 1 \times 0.1 \text{mm}^3$ and were mounted in closed-cycle refrigerators allowing us to investigate a temperature range $15 \text{ K} \leq T \leq 300 \text{ K}$. The components ($Q_h$, $Q_k$, $Q_l$) of the scattering vector are expressed in reciprocal lattice units (r.l.u.) ($Q_h$, $Q_k$, $Q_l$) = ($h*2\pi/a$, $k*2\pi/b$, $l*2\pi/c$) with the lattice constants $a = b \approx 3.97$ Å and $c = 13$ Å of the tetragonal unit cell present at room temperature for all investigated compounds. The energy resolution was 1.5 meV with an incident energy of 23.78 keV. For high momentum resolution the aperture of the analyser was closed to a diameter of 12 mm which covers $0.076°$ in the scattering angle $2\theta$ or $0.01 \times 2\pi/a$ in reciprocal space. Phonon excitations measured as a function of energy transfer at a fixed momentum transfer (constant $Q$-scans) were approximated by damped harmonic oscillator (DHO) functions [18].

## III. Results

All IXS measurements were performed in the Brillouin zone adjacent to $\tau = (4,0,0)$ at phonon wavevectors $q = (0, \pm 0.075, 0)$ and data at selected temperatures are shown in Fig. 1(a) for $Ba_{0.67}K_{0.33}Fe_2As_2$. The wavevector was chosen as the one closest to the zone center where we can still distinguish between the TA phonon and the elastic line. Here, the TA phonon energy at room temperature is 1.96 meV. Upon cooling to T = 35 K ($\approx T_c$), there is a clear softening followed by a slight hardening on further cooling in the superconducting state to T = 15 K. The small absolute size of the effects ($\Delta E_{300K-35K} = 0.25$ meV) required high statistics for an accurate determination of the peak position. In order to compensate for any shift in the elastic energy position[a] we always measured the Stokes and anti-Stokes line of the TA phonon. Since the two phonon peaks appear at $\pm E_{phon}$ we can very accurately determine the elastic position by approximating both peaks with a single DHO function. Since the anti-Stokes line becomes very weak for T ≤ 10 K ($k_B \times 10$ K = 0.862 meV << $E_{phon}$), we could only use this procedure for accurate determination of the excitation energies down to T = 15 K. Furthermore, at each temperature we performed scans on both sides of the (4,0,0) zone center at $Q = \tau + q$ = (4,±0.075,0) in order to compensate for any non-perfect sample orientation which might result in different phonon energies observed at the two wavevectors. We note that the line widths of the approximated DHO functions did not reveal any indication for significant electron-phonon coupling at low temperatures.

In the long-wave-length limit, the dispersion of the TA phonon mode can be described by a linear relation between the phonon energy $E$ and the phonon wave vector $q$,

$$E(|q|) = \sqrt{C_{66}/\rho} \times |q| \quad (2)$$

with $\rho$ being the mass density [19]. Previous results [13,20] show that the dispersion of the TA mode at room temperature in other iron-based superconductors at wave vectors up to $q = (0,0.075,0)$ can be well approximated by this equation. At low temperatures, non-linear dispersions were observed triggered by a strong increase of the nematic correlation length $\xi$ (see also Eq. 3 and the corresponding discussion). However, our analysis will show that the linear regime extends farther in hole-doped compounds because $\xi$ is shorter and, hence, a linear approximation is sensible for the here presented results. Employing equation (2), we can determine $C_{66}(T)$ from our measurements as illustrated in the inset in Fig. 1(b). Corresponding results for two hole-doped compounds at optimal doping, $Ba_{0.6}Na_{0.4}Fe_2As_2$ and $Ba_{0.67}K_{0.33}Fe_2As_2$, are shown in Fig. 1(b) and (c), respectively. Both these compositions are in the part of the phase diagram were $T_c$ is optimal and the structural-magnetic transitions has been completely suppressed by doping [5,11,14,21]. Superconductivity sets in at $T_c$ = 36 K and 34 K in $Ba_{0.67}K_{0.33}Fe_2As_2$ and $Ba_{0.6}Na_{0.4}Fe_2As_2$, respectively (see table I for all transition temperatures in the investigated samples). The softening from room temperature down to $T_c$ is clearly apparent. On further cooling we observe a slight hardening in the superconducting phase for both compounds.

We compare our data to results for the Young's modulus $Y_{[110]}$ obtained via a three-point-bending setup in a capacitance dilatometer, where it has been shown that the temperature dependence of $Y_{[110]}$ is dominated by that of $C_{66}$ if the latter becomes small/soft [11,22]. Hence, it has been used to study nematic fluctuations and determine the nematic susceptibility of various iron-based superconductors [11,14,22]. We note that ultrasound experiments to study the elastic constants are not feasible for the hole-doped members of the 122 family studied here, since such experiments require samples of considerable size and quality [22]. It is however well-known that hole-doped compounds grow only in small and thin platelets, which has also been a considerable problem for inelastic neutron scattering measurements on these systems.

---
[a] Small temperature changes in the monochromator and/or analyzer crystals can result in a drift of the zero energy transfer position.

The IXS data presented in Figure 1 were measured on a sample from the same growth batch of $Ba_{0.67}K_{0.33}Fe_2As_2$ and even the very same sample of $Ba_{0.6}Na_{0.4}Fe_2As_2$ as the measurements of the Young's modulus $Y_{[110]}$ reported in Refs. [11] and [14], respectively. Hence, a comparison is straight forward. While the Young's moduli so far reported were given in a relative scale $Y_{[110]}/Y_{[110]}(T=300K)$ we report $C_{66}$ in an absolute scale in units of GPa [symbols in Fig. 1(b,c)]. We therefore scaled the reported temperature dependences of $Y_{[110]}/Y_{[110]}(T=300K)$ to match the room temperature value of our data [solid lines in Figs. 1(b,c)]. We find that the softening observed in our IXS data is in qualitative agreement with that reported for $Y_{[110]}$: $C_{66}$ softens on cooling down to the superconducting transition temperature but shows a hardening on further cooling. Quantitatively, the softening of $C_{66}$ in $Ba_{0.6}Na_{0.4}Fe_2As_2$ deduced from IXS data (22%) and three-point bending experiments (20%) [14] show a reasonable agreement [Fig. 1(b)].

In $Ba_{0.67}K_{0.33}Fe_2As_2$, $C_{66}$ deduced from IXS data softens by about 25% between room temperature and $T_c$ [dots in Fig. 1(c)] while a maximum softening of only 12% was reported for the Young's modulus $Y_{[110]}$ [11]. Hence, we find that the results for $Y_{[110]}$ yield $C_{66,Y_{[110]}} = 36.5$ GPa for $T = 35$ K [solid line in Fig. 1(c)], whereas the value deduced from IXS at the same temperature is $C_{66,IXS} = 31.9$ GPa. We argue that a scenario in which a $q = 0$ technique, such as the three-point bending setup, reports a $C_{66}$ value which is larger than the one deduced from our IXS experiments is unreasonable and illustrate the situation for the case of $T = 35$ K in the inset of Figure 1(c): The solid line denotes the linear dispersion of the TA phonon using $C_{66,Y_{[110]}} = 36.5$ GPa. It extrapolates to a phonon energy of 1.83 meV at $Q = (4,0.075,0)$. For the same wave vector we determined a phonon energy of $1.7 \pm 0.05$ meV via IXS [dot in the inset of Fig. 1(c)]. The corresponding value $C_{66,IXS} = 31.9$ GPa was used to plot the linear dispersion going through the observed phonon dispersion point [dashed line in the inset of Fig. 1(c)]. Our model presented in [13] can only explain a positive offset of the phonon energies at finite wave vectors compared to the linear extrapolation of the $q = 0$ results [see also Eq. (3) below]. This is corroborated in a recent work by some of us in $Ba(Fe_{0.97}Co_{0.03})_2As_2$ and FeSe [20]. The analysis in [20] uses a sinusoidal bare dispersion, which is better suited for the larger wave vector range, $q \leq (0,0.3,0)$, investigated. However, it also demonstrates that deviations from the linear approximation used in the presented analysis on hole-doped compounds are negligible at $q = (0,0.075,0)$. Thus, we believe that the difference in the softening between the Young's modulus $Y_{[110]}$ [11] and our results from phonon spectroscopy for $Ba_{0.67}K_{0.33}Fe_2As_2$ is not an intrinsic effect but due to uncertainties in the experimental setup. For example, experiments in a three-point bending setup require some pressure on the sample. Furthermore, $C_{66}$ only dominates the behavior of $Y_{[110]}$ if it is small compared to other elastic moduli [11,22]. However, while there is a discrepancy in the quantitative values obtained with these different probes, we obtain good qualitative agreement by scaling the $C_{66}$ values obtained from the three point measurements, $Y_{[110]}$, at two temperatures, $T = 300$ K and $T = 35$ K, which leads to the dashed blue line in Fig. 1 (c).

In fact, the qualitative agreement between $Y_{[110]}$ and the $C_{66}$ values deduced from IXS for both optimally doped compounds is somewhat surprising since clear differences were observed in the optimally electron-doped $Ba(Fe_{0.94}Co_{0.06})_2As_2$ [13]. Using inelastic neutron scattering - large single crystals are available for $Ba(Fe_{0.94}Co_{0.06})_2As_2$ -, some of us reported a qualitative difference between the softening of the TA phonon mode at wavevectors $\boldsymbol{q} = (0, k,0)$, with k ranging from 0.05 to 0.1, and results for $C_{66}$ and $Y_{[110]}$ obtained by resonant ultrasound [12] and three-point-bending experiments, respectively [11]. Whereas the latter show a sharp minimum of $C_{66}$ and $Y_{[110]}$ at the superconducting transition temperature $T_c$, the phonons initially softened upon cooling but then the softening levelled off well above $T_c$ and the phonon energy remained constant down to $T_c$. Upon further cooling into the superconducting phase, the phonons were observed to soften again, opposite to the observations of $C_{66}$ and $Y_{[110]}$. This discrepancy was explained by the presence of fluctuating nematic correlations in the tetragonal state with a finite correlation length $\xi$. The presence of such nematic fluctuations leads to a correction to the normally linear dispersion of the TA phonon mode at small wavevectors given by [13]

$$E(q) = \sqrt{\frac{C_{66}^0}{\rho\left(1+\frac{C_{66}^0/C_{66}-1}{1+\xi^2 q^2}\right)}q^2} \qquad (3)$$

where $C_{66}^0$ is the shear modulus in absence of nematic fluctuations and $C_{66}$ the value observed at $q = 0$. For this model we expect deviations from a linearly extrapolated dispersion based on $C_{66}$ at wavevectors corresponding to the inverse of $\xi$. When such deviations are visible at $k = 0.05$ r.l.u. as in $Ba(Fe_{0.94}Co_{0.06})_2As_2$ [13] the correlation length $\xi$ is $\geq 80$ Å $[= (0.05 \text{ r.l.u.})^{-1}]$. Reversely, the fact that we do not see a qualitative difference between the phonon softening observed in IXS and $Y_{[110]}(T)$ in $Ba_{0.6}Na_{0.4}Fe_2As_2$ [Fig. 1(b)] and $Ba_{0.67}K_{0.33}Fe_2As_2$ [Fig. 1(c)] indicates that the nematic correlation length in these materials is significantly smaller than 53 Å $[= (0.075 \text{ r.l.u.})^{-1}]$. We cannot provide a more quantitative determination of the nematic correlation length since the uncertainties in the absolute value of $C_{66}$ based on $Y_{[110]}$ (see discussion above) prohibit to use it as an input parameter for the initial slope of the phonon dispersion at $q = 0$.

A distinct advantage using phonon spectroscopy to determine the soft elastic constant in 122 iron-based superconductors over volume probes is the ability to follow its evolution into the orthorhombic phases and to the SDW order present, e.g. in underdoped 122 compounds. This is not possible in ultrasound nor in three-point-bending experiments since the unavoidable formation of orthorhombic domains makes an analysis of the obtained data impossible as explained in detail in [22].

However, the soft elastic constant that we measure in the orthorhombic phase is no longer $C_{66}$. The $a$ and $b$ axes of the orthorhombic unit cell are rotated by about 45° with respect to the tetragonal ones and the $(4,0,0)_t$ Bragg peak corresponds to $(4,4,0)_o$ in the orthorhombic structure. Hence, our measurements at temperatures featuring orthorhombic structures (orange shaded in Fig. 3) probe approximately the TA phonon propagating along the $[1\bar{1}0]_o$ direction with polarization $[110]_o$. This is only approximate because we kept a strict transverse scattering geometry, i.e. used a phonon scattering wave vector $Q_{phon} = (4,4,0)_o + q$ with $q \perp (4,4,0)_o$ whereas $[110]_o$ is not perfectly perpendicular to $[1\bar{1}0]_o$ for an orthorhombic structure with $a \neq b$. The reported values of the orthorhombic splitting $\delta = (a-b)/(a+b)$ for the here investigated compounds $Sr_{0.67}Na_{0.33}Fe_2As_2$ ($\delta = 0.003$) [23], $Ba_{0.735}Na_{0.265}Fe_2As_2$ ($\delta = 0.001$) [14] and $Ba_{0.78}K_{0.22}Fe_2As_2$ ($\delta = 0.0025$) [4] correspond to an angular offset between $q$ and $[1\bar{1}0]_o$ of 0.68°, 0.22° and 0.57°, respectively. We note that these are the largest values of $\delta$ in the respective compounds. At temperatures close to the phase transition, the orthorhombic distortion is much smaller and the offset is negligible.

For the case of a negligible small orthorhombic splitting, i.e. $a \approx b$, the linear dispersion at small wave vectors of the TA phonon mode propagating along the $[110]_o$ direction is given by $E(q) = \sqrt{C_o/\rho} \times |q|$, with the elastic constant $C_0 = [C_{11} + C_{22} + 2C_{66} - \sqrt{(C_{11} - C_{22})^2 + 4(C_{12} + C_{66})^2}]/4$, where all components $C_{ij}$ refer to the orthorhombic elastic tensor. In the general case $a \neq b$, the respective elastic constant is a more complex function of $C_{ij}$ displaying also a temperature dependence according to that of the orthorhombic splitting $\delta$. For simplicity in Figure 3, we use only the label $C_{66}$ for the vertical axes and include explanations in the caption. Further, we will continue to refer to the soft elastic constant as $C_{66}$ of the tetragonal elastic tensor for a better readability.

Here, we use IXS to investigate the evolution of $C_{66}$ in $Sr_{0.67}Na_{0.33}Fe_2As_2$, $Ba_{0.735}Na_{0.265}Fe_2As_2$ and $Ba_{0.78}K_{0.22}Fe_2As_2$. While all three compounds feature a structural-magnetic phase transition at $T_{s,N}$, the former two show an additional reentrant tetragonal but magnetically ordered phase below a temperature $T_1 < T_{s,N}$ [5,14,21]. Furthermore, all samples are superconducting at low temperatures. The corresponding superconducting transition temperatures were, however, too low for our investigation, i.e. $T_c < 15$ K. All transition temperatures are listed in table 1. Transition temperatures were obtained by thermal expansion or three-point-bending experiments [11,14,24].

While IXS measurements in twinned samples are possible, phonon dispersions will emanate from the Bragg peaks of each domain, which are at slightly different spots in reciprocal space. Thus, phonon peaks observed via IXS will be broadened in energy, hampering an accurate determination of the excitation energy. The twinning is illustrated by the scan taken at zero energy transfer across the tetragonal (4,0,0) peak along the transverse direction [Fig. 2(a)]. In the tetragonal phase, there is one peak (T=130 K, squares), as expected for a single phase crystal, but two additional peaks appear on either side in the orthorhombic phase indicating the presence of multiple domains (T=100 K, dots). However, by repeated cycling across $T_{s,N}$ (2-3 times), we were able to completely suppress the satellites, i.e., probing only a single domain with the focused x-ray beam, as is demonstrated by the presence of only a single peak in the orthorhombic phase (T=100 K, triangles). This complete suppression of satellites was successful for all three samples featuring an orthorhombic phase, indicating that the orthorhombic domains in all cases were larger than our beam cross section of about 20 µm × 30 µm.

A particular interesting case are the samples featuring the reentrant tetragonal phase below $T_1$. The IXS data for $Sr_{0.67}Na_{0.33}Fe_2As_2$ reflect the two structural phase transitions [Fig. 2(b,c)]. The initial softening [Fig. 2(b)] is followed by a hardening below 120 K in the orthorhombic phase. The onset of the reentrant tetragonal phase results in a very sharp softening on cooling from 60 K to 50 K whereas further cooling leads to a small hardening [Fig. 2(c)]. The corresponding $C_{66}$ values deduced from our IXS data clearly show the two structural transitions [Fig 3(a)]. Of interest is the observation of the hardening of $C_{66}$ when entering the orthorhombic phase, which has not yet been reported in these samples. Upon further cooling, we observe a sudden drop in the phonon energy at $T_1$, indicative of the 1$^{st}$ order character of this transition from the orthorhombic to the reentrant tetragonal state [25]. It is intriguing that $C_{66}$ is softest in the reentrant tetragonal phase. Unfortunately, comparative data for $Y_{[110]}$ in $Sr_{0.67}Na_{0.33}Fe_2As_2$ are not available/published. However, temperature dependent neutron pair-distribution function analysis found local orthorhombic fluctuations on the length scale of 20 Å in the $C_4$ phase of $Sr_{1-x}Na_xFe_2As_2$ [26] indicating a large nematic susceptibility in agreement with our results.

A rather soft shear modulus in the $C_4$ phase was recently reported for $Ba_{0.735}Na_{0.265}Fe_2As_2$ by experiments in a three-point bending setup [solid line in Fig. 3(b)] [14] and is corroborated by our results [symbols in Fig. 3(b)] obtained for the very same sample as used in [14]. However, we neither see the pronounced decrease just above the transition into the orthorhombic phase at $T_{s,N}$ nor the strong recovery after entering the low-temperature tetragonal phase at $T_1$. Going back to phonon dispersions, the differences between $C_{66}$ values deduced from IXS and those based on $Y_{[110]}$ near $T_{s,N}$ and $T_1$ [Fig. 3(b)] imply that the finite wave vector phonon is observed by IXS at a higher energy than it is expected based on a linear extrapolation of the $C_{66}$ value determined at $q = 0$ by $Y_{[110]}$. Within our model dispersion (Eq. 3) such a positive offset of the phonon energy from the $q=0$ linear extrapolation would indicate that the nematic correlation length $\xi$ is similar or larger than the inverse of the investigated wave vector. A large correlation length seems to be plausible near the phase transition temperatures in an under-doped compound. However, similar to the situation for optimally doped compounds, we cannot provide a quantitative determination of the nematic correlation length since the uncertainties in the absolute value of $C_{66}$ based on $Y_{[110]}$ prohibit to use it as an input parameter for the initial slope of the phonon dispersion at $q = 0$.

To complement our report, we show our IXS results for under-doped $Ba_{0.78}K_{0.22}Fe_2As_2$ [Fig. 3(c)]. The hardening below $T_{s,N}$ is the strongest observed for all three samples investigated. For this sample, $T_c = 24.5$ K is high enough that we were able to accurately determine the phonon energy and determine $C_{66}$ through the superconducting transition. Since strong competition between superconductivity and orthorhombicity is well established, one might expect a softening upon entering the superconducting phase. But we do not see any significant effects outside the statistical error of our measurement.

The discrepancy between the IXSs results and measurements of the shear modulus via three-point-bending setups can be understood both quantitatively and qualitatively as resulting from the inability of three-point-bending setups to deduce sensible results for investigating orthorhombic samples. The shear modulus $C_{66}$ is in fact expected to show a clear hardening below the nematic/structural phase transition within mean-field theory [as shown in the inset in Fig. 3(c)] [22] which is however neither the case for the measured $Y_{[110]}$ [blue lines in Figs. 3(b,c)] nor the ultrasound measurements, e.g., in underdoped $Ba(Fe_{1-x}Co_x)_2As_2$ [12]. Hence, the qualitative temperature dependence of $C_{66}$ deduced from IXS reflects the real physics with a hardening on cooling below the nematic/structural phase transition.

Additional quantitative differences, e.g., in optimally doped $Ba(Fe_{0.67}K_{0.33})_2As_2$ [Fig. 1(c)] may arise because of the fact that $Y_{[110]}(T)$ only reflects $C_{66}(T)$ if the latter is very small, e.g., close to the nematic transition. When this transition is suppressed the softening of $Y_{[110]}$ is much reduced and, hence, $Y_{[110]}$ may differ more strongly from $C_{66}$. Furthermore, IXS is performed at finite wavevectors. For a mean-field-like second-order structural phase transition at $q = 0$, we expect that the soft elastic constant and, hence the slope of the soft phonon mode at the zone center, vanishes at the phase transition temperature. The phonon branch will recover its normal dispersion for wave vectors far enough away from the zone center. Hence, the softening is complete only at the zone center, i.e., $q = 0$ and gradually decreases with increasing value of the wave vector. A instructive example is the structural phase transition which precedes the superconducting one in the $A15$ superconductor $Nb_3Sn$ (for a review see [27]). While ultrasound experiments [28] show a complete softening at $q = 0$, phonon measurements at finite momenta report a maximum softening of about 60% quickly decreasing going away from the zone center.

Lastly, quantitative results from three-point-bending setups depend sensitively on the proper alignment of the sample within the experimental setup, which cannot be checked with such a high precision as available by high resolution Bragg scattering ($\Delta_{angle} < 0.1°$), routinely performed during an IXS investigation. For instance, uncertainties in geometrical parameters of the three-point-bending setup were given as reason that only normalized values of the Young's modulus $Y_{[110]}$ were reported [11]. In contrast, we find that elastic constants deduced from IXS provides physically reliable results also in the orthorhombic structure of 122 iron-based superconductors.

### IV. Discussion

We now turn to discuss our results in the wider context of nematic correlations. Within the spin-driven nematic scenario there is a direct link between the nematic correlation length $\xi$ and the magnetic correlation length $\xi_{AFM}$. In particular, $\xi$ diverges when $\xi_{AFM}$ reaches a certain threshold value [29]. Thus, the increase of $\xi$ in $Ba(Fe_{0.94}Co_{0.06})_2As_2$ on cooling towards and the subsequent decrease below $T_c$ were rationalized via the strong competition between magnetism and superconductivity [13]. Our results for optimally hole-doped $Ba_{0.67}K_{0.33}Fe_2As_2$ and $Ba_{0.6}Na_{0.4}Fe_2As_2$ do not provide any indication for a significant increase of $\xi$ and, thus, indicate a rather small magnetic correlation length. We note that there is also a clear discrepancy between the nematic susceptibilities $\chi_{nem}$ deduced from three-point bending experiments for $Ba(Fe_{0.94}Co_{0.06})_2As_2$ on the one side and $Ba_{0.67}K_{0.33}Fe_2As_2$ and $Ba_{0.6}Na_{0.4}Fe_2As_2$ on the other side. Whereas $\chi_{nem}$ strongly increases on cooling towards $T_c$ in $Ba(Fe_{0.94}Co_{0.06})_2As_2$ following a Curie-Weiss law [11], the temperature dependence of $\chi_{nem}$ in $Ba_{0.67}K_{0.33}Fe_2As_2$ [11] and $Ba_{0.6}Na_{0.4}Fe_2As_2$ [14] is less pronounced and not diverging close to $T_c$. Yet, the differences is of a qualitative nature since the relative phonon softening observed in the hole-doped compounds, 22% - 25% (see table 1), is similar to the softening reported by ultrasound for optimally doped Co doped $Ba(Fe_{1-x}Co_x)_2As_2$ [12]. Regarding the significantly higher superconducting transition temperatures in $Ba_{0.67}K_{0.33}Fe_2As_2$ ($T_c = 36$ K) and $Ba_{0.6}Na_{0.4}Fe_2As_2$ ($T_c = 34$ K) compared to $Ba(Fe_{0.94}Co_{0.06})_2As_2$ ($T_c = 26$ K [13]), our results indicate that strong nematic and magnetic correlations are detrimental to large values of $T_c$.

Whereas all investigated underdoped samples have coincident nematic/structural and magnetic transitions at $T_{s,N}$, they differ in that $Ba_{0.78}K_{0.22}Fe_2As_2$ does not feature a magnetic tetragonal $C_4$ phase [4]. Furthermore, $Ba_{0.78}K_{0.22}Fe_2As_2$ features a Curie-Weiss like, i.e. divergent nematic susceptibility [11] whereas $\chi_{nem}$ in $Ba_{0.735}Na_{0.265}Fe_2As_2$ features a much less pronounced increase close to $T_{s,N}$ [14] qualitatively similar to the observation in the optimally hole-doped samples discussed above.

The hardening of the soft elastic constants on cooling in the orthorhombic phases below $T_{s,N}$ is very different in the investigated samples. However, all observed hardenings are smaller than in the parent compounds $BaFe_2As_2$ and $SrFe_2As_2$ in which the same phonon's energy increases in the orthorhombic phase to 7% and 17% above its room temperature value, respectively [30]. In $Ba_{0.78}K_{0.22}Fe_2As_2$, the soft elastic constant at low temperatures recovers to more than 90% of the room temperature value [Fig. 3(c)]. In contrast, there is practically no hardening according to our phonon data in the orthorhombic phase of $Ba_{0.735}Na_{0.265}Fe_2As_2$. The soft elastic constant remains around 67% of the room temperature value and displays a small but sudden drop to 62% of the room temperature value on entering the $C_4$ phase [Fig. 3(b)] in agreement with reported strong nematic fluctuations by 3-point bending experiments [14]. Regarding the reported behavior of $\chi_{nem}$ for these compounds [11,14], the hardening of the elastic constant below $T_{s,N}$ seems to reflect the strength of the nematic fluctuations above $T_{s,N}$, which makes sense in that the hardening certainly reflects the stability of the orthorhombic phase.

In $Sr_{0.67}Na_{0.33}Fe_2As_2$ the soft elastic constant hardens to about 79% of its room temperature value in the orthorhombic phase [Fig. 3(a)] and, hence, the hardening is intermediate between the observations for the other two under-doped samples. Note that most of the hardening happens in a rather narrow temperature range of ΔT = 22 K below $T_{s,N}$. Therefore, we argue that nematic fluctuations at $T_{s,N}$ are stronger in $Sr_{0.67}Na_{0.33}Fe_2As_2$ than in $Ba_{0.735}Na_{0.265}Fe_2As_2$. Correspondingly, the sudden drop on entering the $C_4$ phase in $Sr_{0.67}Na_{0.33}Fe_2As_2$ is much more pronounced and the average value of the soft elastic constants in the $C_4$ phase is only 57% of the room temperature value, although the absolute values are similar in both compounds featuring $C_4$ phases. We note that the observed softening and, hence, the presence of nematic fluctuations, from high temperatures to T = 15 K is significantly stronger in the two $C_4$ compounds than in the samples at optimal doping (see table 1). In contrast, the values of $T_c$ are more than three times smaller. Therefore, our conclusion for optimally doped samples that strong nematic fluctuations are detrimental to the absolute value of $T_c$ extend to the $C_4$ phases of $Sr_{1-x}Na_xFe_2As_2$ and $Ba_{1-x}Na_xFe_2As_2$ and are also visible in the sudden drop of $T_c$ on crossing from the $C_2$ to the $C_4$ magnetic phases in the respective phase diagrams [23,24]. Based on the published phase diagram of $Ba_{1-x}K_xFe_2As_2$ [4], we expect strong nematic fluctuations in the $C_4$ phase of this hole-doped compound as well.

## V. Conclusion

We have reported phonon measurements of the in-plane polarized TA phonon in $Ba_{1-x}(K,Na)_xFe_2As_2$ and $Sr_{1-x}Na_xFe_2As_2$ for several different compositions. In optimally doped compounds ($Ba_{0.67}K_{0.33}Fe_2As_2$, $Ba_{0.6}Na_{0.4}Fe_2As_2$), that do not exhibit any structural-magnetic phase transitions, the deduced values of the soft elastic constant $C_{66}$ are in qualitative agreement with previous reports based on a three-point-bending setup in a capacitance dilatometer [11,14,22]. This is in contrast to a recent report on optimally doped $Ba(Fe_{0.94}Co_{0.06})_2As_2$ [13] and shows that the correlation length of nematic fluctuations ξ in optimally hole-doped compounds ($\xi < 53$ Å) is significantly shorter than in the Co-doped system ($\xi \approx 100$ Å). In underdoped $Ba_{0.735}Na_{0.265}Fe_2As_2$ and $Sr_{0.67}Na_{0.33}Fe_2As_2$ we find the softest elastic constants in the magnetic-tetragonal $C_4$ phase. We also report the expected hardening of the soft elastic constant in the orthorhombic phase of underdoped samples not observed in previous investigations. More generally, our results demonstrate that phonon spectroscopy via IXS can determine elastic constants with high accuracy on absolute scales even in orthorhombic and twinned phases.

**Acknowledgements:**
M.K. and F.W. were supported by the Helmholtz Society under contract VH-NG-840. S.R. was supported by the Materials Sciences and Engineering Division, Office of Basic Energy Sciences, U.S. Department of Energy. This research used resources of the Advanced Photon Source, a U.S. Department of Energy (DOE) Office of Science User Facility operated for the DOE Office of Science by Argonne National Laboratory under Contract No. DE-AC02-06CH11357.

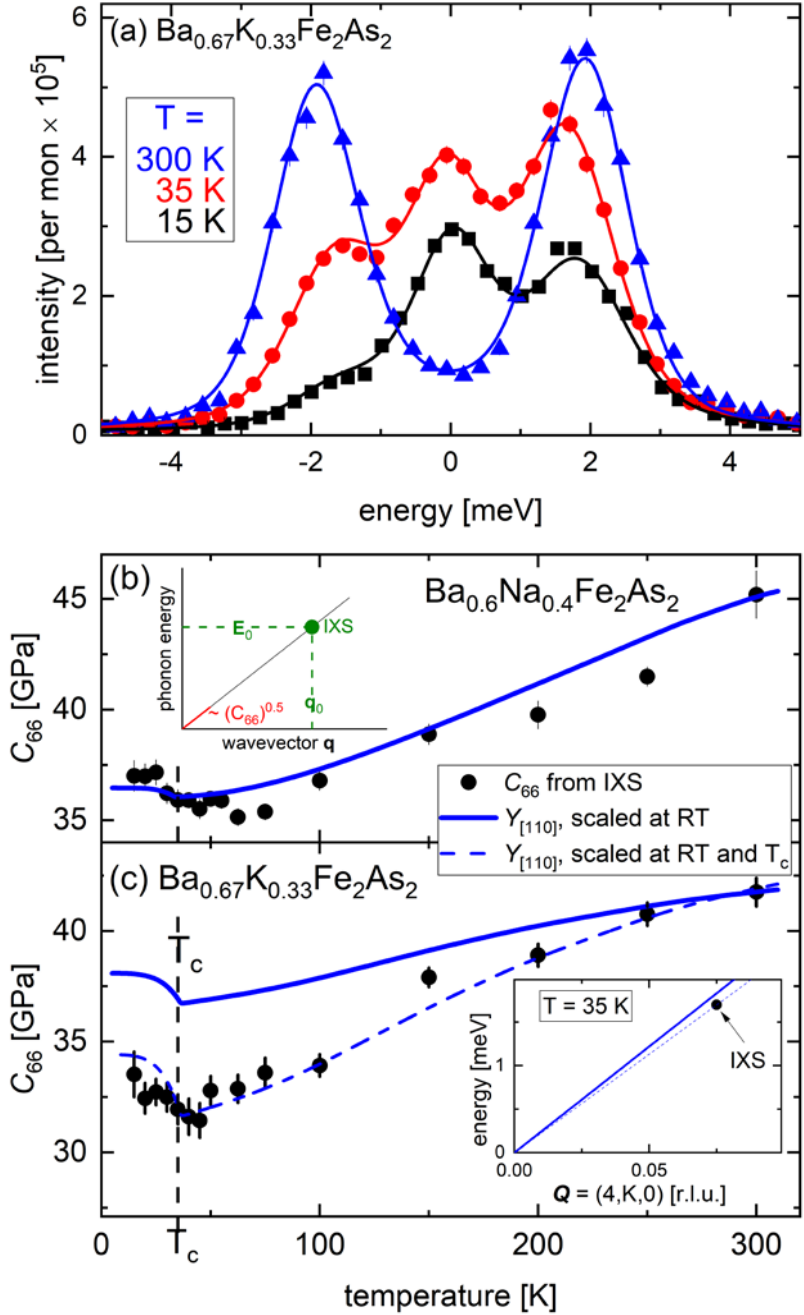

**FIG. 1.** (a) Typical set of raw IXS data for $Ba_{0.67}K_{0.33}Fe_2As_2$. Data at temperatures above, close to and below the superconducting transition temperature $T_c$ = 36 K were taken at symmetry-equivalent wavevectors $Q = (4, \pm 0.075, 0)$ and combined to minimize statistical error bars. The fits (solid lines) consist of a damped-harmonic oscillator (DHO) function for the phonon excitation, a resolution-limited peak for the incoherent elastic scattering and an estimated constant experimental background. The intensities observed at room temperature are down-scaled by a factor of three for a better comparison. (b,c) Shear modulus $C_{66}(T)$ for (b) $Ba_{0.6}Na_{0.4}Fe_2As_2$ (dots) and (c) $Ba_{0.67}K_{0.33}Fe_2As_2$ (dots) deduced from linear dispersion fits to the observed phonon energies [see inset in (b) and text]. Lines represent published results for the Young's modulus $Y_{[110]}$ [11,14] closely related to $C_{66}$ in these compounds. The reported relative temperature dependence of $Y_{[110]}(T)/ Y_{[110]}(T=300K)$ was scaled to match the IXS results at room temperature (RT), i.e. T = 300 K (solid lines). The dashed line in (c) was scaled at two temperatures, i.e. to match the IXS results at room temperature and at $T_c$. The inset in (c) illustrates - for the case of T = 35 K - the scenario that the expected linear dispersion based on $Y_{[110]}$ scaled only at room temperature (solid line) extrapolates to a phonon energy of 1.83 meV at $Q = (4, 0.075, 0)$, which is higher than the one observed in IXS at this wave vector, i.e., 1.7±0.05 meV (dot). The dashed line denotes the linear dispersion based on $Y_{[110]}$ scaled at room temperature and $T_c$. The error bar of the phonon energy (dot) is smaller than the symbol size.

| | $Sr_{0.67}Na_{0.33}Fe_2As_2$ | $Ba_{0.735}Na_{0.265}Fe_2As_2$ | $Ba_{0.78}K_{0.22}Fe_2As_2$ | $Ba_{0.67}K_{0.33}Fe_2As_2$ | $Ba_{0.6}Na_{0.4}Fe_2As_2$ |
|---|---|---|---|---|---|
| $T_{s,N}$ | 122 K | 71 K | 95 | --- | --- |
| $T_1$ | 50 K | 45 K | --- | --- | --- |
| $T_c$ | 10.5 K | 8.5 K | 24.5 | 36 K | 34 K |
| $E_{phon}$ | 2.00(1) meV @ T=300 K | 1.93(2) meV @ T=250 K | 1.93(2) meV @ T=300 K | 1.96(2) meV @ T=300 K | 2.06(2) meV @ T=300 K |
| $\Delta C_{66}$(IXS) | 45% | 38% | 30% | 25% | 22% |

**Table 1.** Transition temperatures for the samples used in our investigation: $T_{s,N}$ refers to the simultaneous structural-magnetic transition at the onset of SDW order. $T_1$ marks the onset of the tetragonal magnetic phase. $T_c$ denotes the superconducting transition temperature. $E_{phon}$ gives the phonon energy at the given temperature. Error bars are given in brackets for the last digit. The last line reports the observed relative softening $\Delta C_{66} = C_{66,min}/C_{66,300K}$ observed via IXS. For $Ba_{0.735}Na_{0.265}Fe_2As_2$, we compare the values at minimum temperature of T = 15K to those at room temperature.

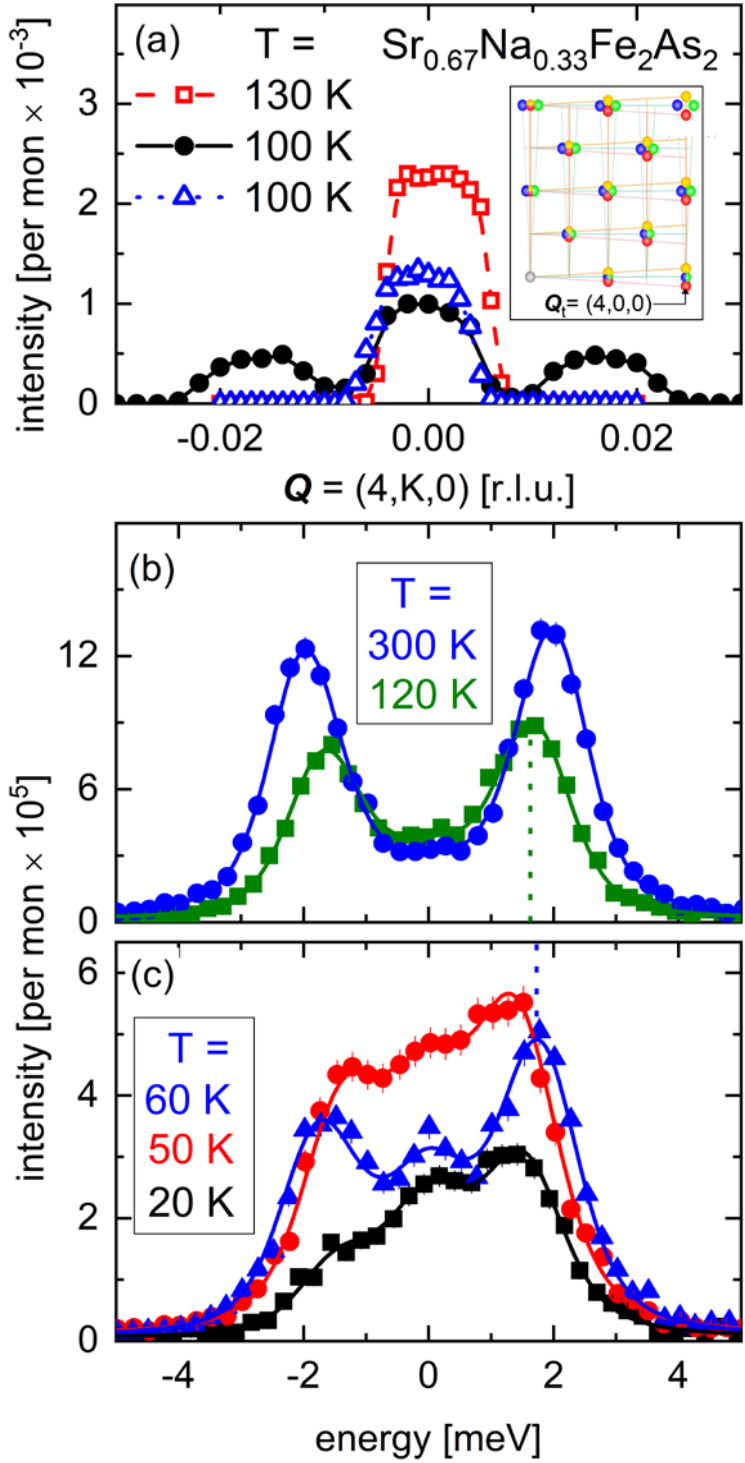

**FIG. 2.** (a) Constant energy scans at E = 0 and $\mathbf{Q}$ = (4, K, 0) in the paramagnetic phase (T = 130 K) and in the orthorhombic phase ($T_1 < T = 100\,K < T_{s,N}$) of $Sr_{0.67}Na_{0.33}Fe_2As_2$. The data demonstrate that we were able to make measurements in a single domain by repeated cooling across $T_{s,N}$. The inset shows the expected splitting of the $(4,0,0)_t$ Bragg reflex in the twinned orthorhombic structure [31]. (b,c) IXS raw data for $Sr_{0.67}Na_{0.33}Fe_2As_2$ at (b) room temperature, close to the structural phase transitions at $T_{s,N}$ = 120 K and (c) $T_1$ = 52 K, and at the minimum temperature investigated T = 20 K (> $T_c$ = 10.5 K). The fits (solid lines) consist of a damped-harmonic oscillator (DHO) function for the phonon excitation, a resolution-limited peak for the incoherent elastic scattering and an estimated constant experimental background. The vertical dotted lines indicate the phonon energy obtained for (b) T = 120 K and (c) T = 60 K for easy cross-panel comparison.

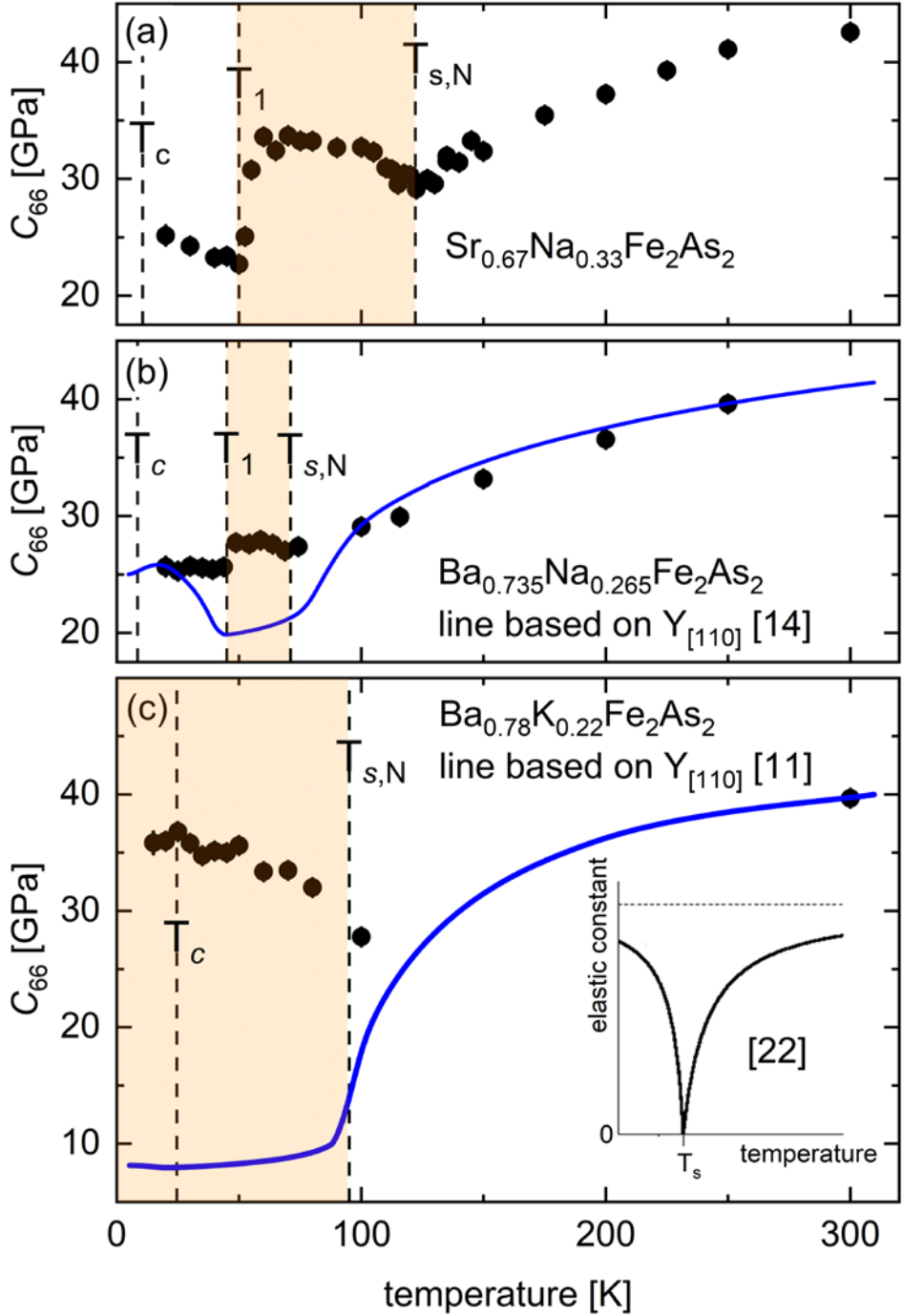

**FIG. 3.** $C_{66}(T)$ as deduced from our phonon measurements in samples featuring full structural phase transitions at $T_{s,N}$: (a) $Sr_{0.67}Na_{0.33}Fe_2As_2$ and (b) $Ba_{0.735}Na_{0.265}Fe_2As_2$ exhibit two structural phase transitions due to the onset of the *C4* magnetic-tetragonal phase at $T_1$, while (c) $Ba_{0.78}K_{0.22}Fe_2As_2$ stays orthorhombic at all temperatures below $T_{s,N}$. Temperatures at which the compounds feature an orthorhombic structure are shaded (orange). The orthorhombic unit cell is rotated by about 45° in the basal plane with respect to the tetragonal one. Hence, we investigate approximately (see text) the TA phonon propagating along the $[110]_o$ direction with $[1\bar{1}0]_o$ polarization in the orthorhombic structure. The elastic constant defining the slope of this mode in the orthorhombic structure is $C_0 = [C_{11} + C_{22} + 2C_{66} - \sqrt{(C_{11} - C_{22})^2 + 4(C_{12} + C_{66})^2}]/4$, where all components $C_{ij}$ refer to the orthorhombic elastic tensor (see text). The lines indicate the reported temperature dependences of the Young's moduli $Y_{[110]}(T)$ [11,14] scaled at (b) T = 250K and (c) T = 300 K. Comparable data for $Sr_{0.67}Na_{0.33}Fe_2As_2$ are not available. The inset in (c) shows the expected mean-field behavior of the soft elastic constant at a nematic phase transition (solid line). The dashed line denotes the unrenormalized elastic constant (dashed horizontal line). Figure taken from [22].